\documentclass[aps,showpacs,pra,amsfonts,superscriptaddress,twocolumn]{revtex4-1}  
\usepackage[english,ngerman]{babel}
\usepackage{graphicx}  
\usepackage{dcolumn}   
\usepackage{bm}        
\usepackage{bbm}
\usepackage{amssymb}   
\usepackage{amsmath}
\usepackage{mathtools}
\usepackage[utf8]{inputenc}
\usepackage{changes}
\usepackage{times}
\usepackage{hyperref}
\usepackage{dsfont}
\usepackage{outlines}

\definecolor{myurlcolor}{rgb}{0,0,0.7}
\definecolor{myrefcolor}{rgb}{0.8,0,0}
\usepackage{hyperref}
\hypersetup{colorlinks, linkcolor=myrefcolor,
citecolor=myurlcolor, urlcolor=myurlcolor}

\usepackage{amsthm}

\usepackage{cleveref}

\newcommand{\ignore}[1]{}

\newcommand{\rt}{\rho^t}
\newcommand{\re}{\rho^e}

\newcommand{\eg}{\emph{e.g. }}			
\newcommand{\ie}{\emph{i.e. }}				

\newcommand{\vect}[1]{{\bm{#1}}}			
\newcommand{\avg}[1]{\langle#1\rangle}		
\newcommand{\bra}[1]{\langle#1|}			
\newcommand{\ket}[1]{|#1\rangle}			

\newcommand{\vecu}{\ensuremath{\vect{u}}}

\begin{document}
\selectlanguage{english}

\title{Bounding the fidelity of quantum many-body states from partial information}

\date{\today}

\author{Matteo Fadel} 
\affiliation{Department of Physics, University of Basel, Klingelbergstrasse 82, 4056 Basel, Switzerland}

\author{Albert Aloy}
\affiliation{ICFO-Institut de Ciencies Fotoniques, The Barcelona Institute of Science and Technology, 08860 Castelldefels (Barcelona), Spain}

\author{Jordi Tura} 
\affiliation{Max-Planck-Institut f\"ur Quantenoptik, Hans-Kopfermann-Stra{\ss}e 1, 85748 Garching, Germany}

\begin{abstract}
We formulate an algorithm to lower bound the fidelity between quantum many-body states only from partial information, such as the one accessible by few-body observables. Our method is especially tailored to permutationally invariant states, but it gives nontrivial results in all situations where this symmetry is even partial. This property makes it particularly useful for experiments with atomic ensembles, where relevant many-body states can be certified from collective measurements.  As an example, we show that a $\xi^2\approx-6\;\text{dB}$ spin squeezed state of $N=100$ particles can be certified with a fidelity up to $F=0.999$, only from the measurement of its polarization and of its squeezed quadrature. Moreover, we show how to quantitatively account for both measurement noise and partial symmetry in the states, which makes our method useful in realistic experimental situations.
\end{abstract}

\maketitle

\section{Introduction} 
Quantum mechanics describes physical states with a finite number of degrees of freedom through density matrices. These are positive semidefinite, normalized operators acting on the Hilbert space associated to the system. The question of how ``close'' two quantum states are arises naturally in multiple situations, and the natural answer is given by the Bures \cite{BuresTAMS1969} (or Helstrom \cite{HelstromPLA1967}) metric, according to which the distance $\mathcal{D}$ between two density matrices is given by $\mathcal{D}(\rho_1,\rho_2) ^2 := 2 \left( 1 - \sqrt{F(\rho_1,\rho_2)} \right)$, where $F(\rho_1,\rho_2)$ is the fidelity function \cite{JozsaJMO1994} defined as
\begin{equation}\label{eq:amatriciana}
F(\rho_1,\rho_2) := \rm{Tr}\left[ \sqrt{ \sqrt{\rho_1} \rho_2 \sqrt{\rho_1} } \right]^2 \;.
\end{equation}

While the fidelity Eq.~\eqref{eq:amatriciana} is not a metric by itself, it constitutes nevertheless a notion of how close two quantum states are. For this reason, it is a key pillar of a vast number of results in quantum information theory, including entanglement quantification \cite{VedralPRL1997} and concentration \cite{BennettPRA1996}, quantum computation \cite{LidarPRL1998} or quantum estimation for metrology \cite{ParisIJQI2009}.

Note that computing the fidelity Eq.~\eqref{eq:amatriciana} requires the full knowledge of the density matrices $\rho_{1,2}$ \cite{ODonnell2015, ODonnell2016, Straupe2016, HaahIEEE2017, CramerNatComms2010}. This constitutes a severe obstacle in experimental situations, where full state tomography is typically impractical and unreliable, especially in the many-body regime \cite{TitchenerNPJ2018, TorlaiNatPhys2018, Granade2017, SchwemmerPRL2015}.

Here, we present a method to lower bound the fidelity between two quantum states when only partial information on few-body marginals is available. Our method is especially tailored to permutationally invariant states, but it gives nontrivial results even when this symmetry is only partial. The algorithm we propose is based on a semidefinite program (SdP) that looks for a density matrix that minimizes the fidelity with respect to the target state, while maintaining compatibility with the available data, and exploiting additional constraints resulting from the quantum marginal problem for symmetric states \cite{Aloy2020}.

We exemplify the applicability of our results in relevant many-body systems, and offer a tool to bound the fidelity of their state that is useful for state characterization \cite{BavarescoNatPhys2018}, entanglement quantification \cite{WoottersPRL1998, BrussJMP2002}, estimation of the Fisher information \cite{AugusiakPRA2016}, state tomography \cite{TothPRL2010}, and other information-theoretic tasks.

\section{Preliminaries}
Consider the task of preparing a target quantum state $\rt$ of $N$ qudits, not necessarily pure, and let $\re$ denote the state actually prepared in the experiment. The natural figure of merit for this task is Eq.~\eqref{eq:amatriciana} \cite{JozsaJMO1994}. Suppose however we do not fully know $\re$, but only the expectation value of $M$ observables $D_i := \mathrm{Tr}[O_i \re]$, with $i = 1\ldots M$. Can we find out what is the worst-case fidelity with the target state, that is compatible with the experimental data?

This question can be formulated as an algorithm, namely
\begin{equation}
  \begin{array}{llll}
  \text{minimize}_{\rho} & F(\rho,\rt) &&\\
  \mbox{such that}& \rho &\succeq & 0\\
 & \rm{Tr}[\rho] &=&1\\
  &\rm{Tr}[O_i \rho] &=& D_i \qquad \forall 1 \leq i \leq M.
 \end{array}
 \label{eq:carbonara}
\end{equation}
In words, Eq.~\eqref{eq:carbonara} looks for a valid density matrix $\rho$ compatible with the data $D_i$, which minimizes $F(\rho,\rt)$. The result is a matrix $\rho^\ast$ such that $F(\rho^\ast,\rt)\leq F(\re,\rt)$, \ie giving a lower bound on the fidelity of $\re$.

In the following, we show that Eq.~\eqref{eq:carbonara} can be casted into an efficient SdP. We focus on the experimentally relevant case where $\re$ is a symmetric state and $O_i$ constitutes a $k$-order moment of a collective observable, \eg $O_i = S^k$ where $S$ is the total spin operator along a given direction. Then, we observe the following: (a) The $k$-body reduced density matrix $\sigma$ of $\re$ ($k$-RDM) suffices to calculate $\avg{S^k}$, (b) when $\rt$ is pure, Eq.~\eqref{eq:amatriciana} becomes linear in $\re$, namely $F(\re,\rt) = \avg{\re,\rt}$, and in general, the concavity of Eq.~\eqref{eq:amatriciana} guarantees $F(\re,\rt) \geq \avg{\re,\rt}$  \cite{JozsaJMO1994}, (c) the compatibility constraints of a $k$-RDM with a global symmetric state can be checked efficiently with an SdP \cite{Aloy2020}. The combination of these properties allow us to give an efficient formulation of Eq.~\eqref{eq:carbonara}.

Let us emphasize that, given the optimization task in Eq.~\eqref{eq:carbonara}, it is not \textit{a priori} clear how many/which observables $O_i$ are needed in order to obtain a nontrivial bound for the fidelity, nor even if all of them up to a given order $m$ are sufficient. To build and intuition towards answering these questions we proceed in two steps: (i) we ask whether the exact knowledge of the full $m$-RDM $\sigma$ is sufficient to give a nontrivial bound; (ii) we relax the requirement of knowing the full $m$-RDM, and study whether only partial information on it is sufficient.

\begin{figure}
  \centering
  \includegraphics[width=0.97\linewidth]{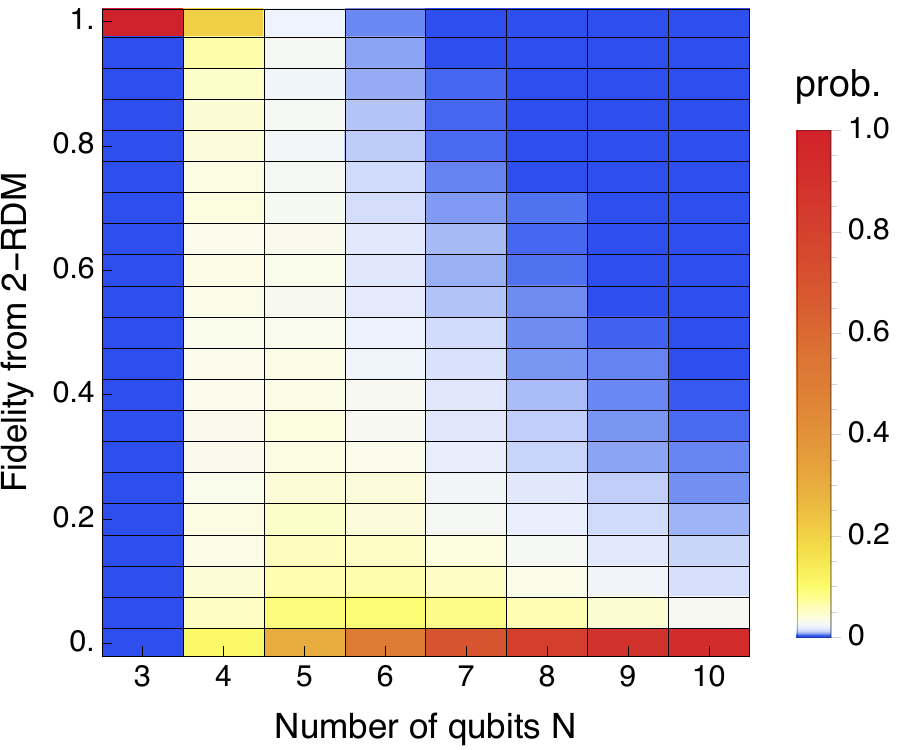}
  \caption{Fidelity bounds given by SdP \eqref{eq:gricia} for random pure $\rt$ and their $2$-RDM $\sigma$. For each $N$ we perform $10^4$ trials and plot the (normalized) histogram of the results as a color scale. One observes that the $2$-RDM of a generic pure $\rt$ rapidly fails to represent it faithfully.}\label{fig:FF2RDM}
\end{figure}

\section{Fidelity from a fully characterized $m$-RDM}
To begin, note that the entries $\sigma^{\boldsymbol{\alpha}}_{\boldsymbol{\beta}}$ of a symmetric $m$-qudit density matrix $\sigma$ are indexed by the partitions of the integer $m$ into $d$ elements, denoted $\boldsymbol{\alpha}, \boldsymbol{\beta} \vdash m$. Furthermore, the $m$-RDM $\sigma$ of a symmetric state $\rho$ can be efficiently computed as a linear function $\langle A^{\boldsymbol \alpha}_{\boldsymbol{\beta}}, \rho \rangle = \sigma^{\boldsymbol{\alpha}}_{\boldsymbol{\beta}}$ of the coefficients of $\rho$, where the matrices $A^{\boldsymbol \alpha}_{\boldsymbol{\beta}}$ stem from the partial trace in the symmetric space \cite{Aloy2020}. Full characterization of $\sigma$ means by (a) that $M$ is sufficiently large so that $O_i$ may form an informationally complete set of observables on $\sigma$. Hence, allowed by (c), we can formulate the following SdP
\begin{equation}
  \begin{array}{llll}
  \min_{\rho} & \avg{\rho,\rt} &&\\
  \mbox{s.t.}& \rho &\succeq & 0\\
  &\langle A^{\boldsymbol \alpha}_{\boldsymbol{\beta}}, \rho \rangle &=& \sigma^{\boldsymbol{\alpha}}_{\boldsymbol{\beta}} \qquad \forall \boldsymbol{\alpha}, \boldsymbol{\beta} \vdash m,
 \end{array}
 \label{eq:gricia}
\end{equation}
which certifies a lower bound on the solution of Eq.~\eqref{eq:carbonara}. In addition, when (b) holds, both solutions are equal.

In Fig.~\ref{fig:FF2RDM} we use SdP~\eqref{eq:gricia} to investigate under which generic conditions does a 2-RDM $\sigma$ coming from a pure $N$-qubit $\rt$ uniquely determines $\rt$. To this end, we: (1) Pick a symmetric $\rt$ of a given rank at random; (2) Compute its 2-RDM $\sigma$; (3) use SdP~\eqref{eq:gricia} to find a state $\rho$ that minimizes the fidelity $F(\rho, \rt)$; (4) iterate a large number of times.
If the fidelity is one up to numerical accuracy error, this is an indication that the global state $\rt$ is uniquely identified by its 2-RDM. On the other hand, if the fidelity falls significantly below one, it shows that the 2-RDM is no longer sufficient to faithfully represent the original state, since a global extension to a symmetric state of size $N$ is not unique.

\begin{figure}
  \centering
  \includegraphics[width=0.94\linewidth]{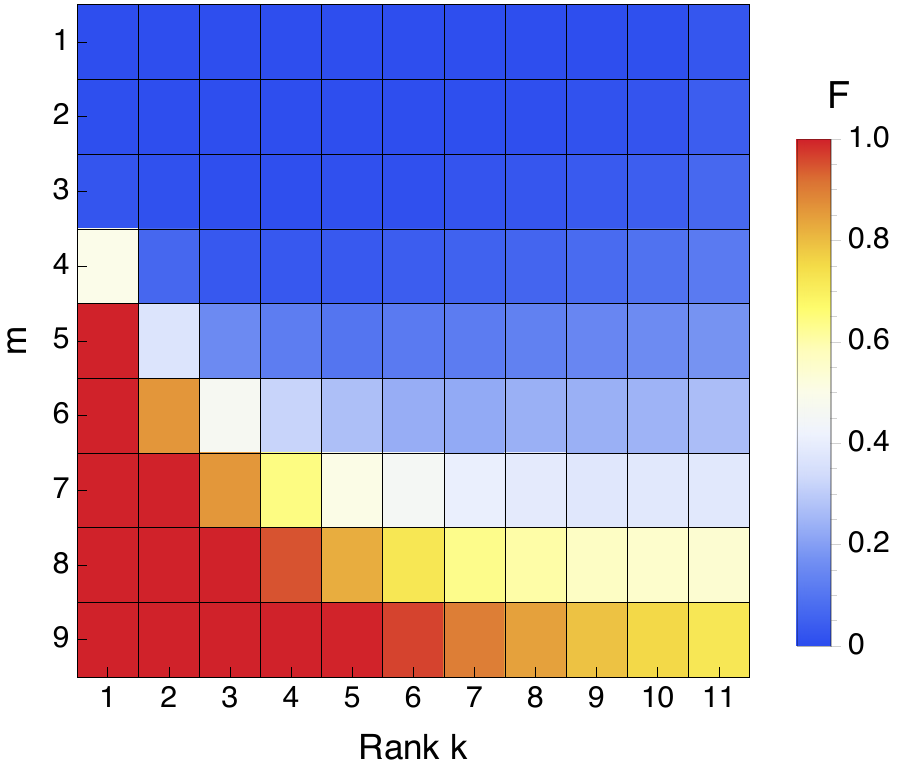}
  \caption{Fidelity bounds given by SdP \eqref{eq:gricia} for random $10$-qubits states with $\text{rank}(\rt)=k$ and their $m$-RDM $\sigma$. For each pixel we plotted the mean fidelity for $10^2$ trials. Red pixels indicate when the $m$-RDM faithfully represent a generic $\rt$ of rank $k$.}\label{fig:AFTn30}
\end{figure}

In more generality, one may be interested in lower-bounding the fidelity of an experimental attempt to prepare a mixed quantum state $\rt$ from a $m$-RDM.
In Fig.~\ref{fig:AFTn30} we show numerical results for random $N=10$ qubits density matrices $\rt$ of rank $k$, illustrating the average lower bound on the fidelity attainable by looking at their $m$-RDM. One can observe a certain correlation between $m$ and $k$, showing more chances to have a unique extension for low $k$ even after tracing out a significant fraction of the particles. 

It is worth emphasizing that SdP~\eqref{eq:gricia} does not rely on any approximation to check the compatibility with a RDM, since the constraints given by $A^{\boldsymbol \alpha}_{\boldsymbol{\beta}}$ are exact. Therefore, we also expect the resulting bound for the fidelity to be always the highest attainable from the available information.

Let us now focus on an experimentally relevant scenario, where the aim is to prepare a $N$-partite pure state ($\text{rank}(\rt)=1$) and we are limited to measuring $O_i = S^{k_i}$ for $k_i\leq m$ and $m$ small.
From our previous results we can conclude that, unfortunately, for a generic state, the lower bound on the fidelity that can be estimated from the $m$-RDM alone rapidly tends to zero as $N$ increases. Hence, in the generic case, it is not always possible to provide non-trivial lower bounds on the fidelity from RDMs.
Nevertheless, this might still be possible for physically relevant classes of states.
Here we focus on spin squeezed states, an important class which is routinely prepared in experiments, and used for a variety of applications ranging from quantum metrology to tests of fundamental quantum physics \cite{PezzeRMP2018}.

\begin{figure}
  \centering
  \includegraphics[width=0.93\linewidth]{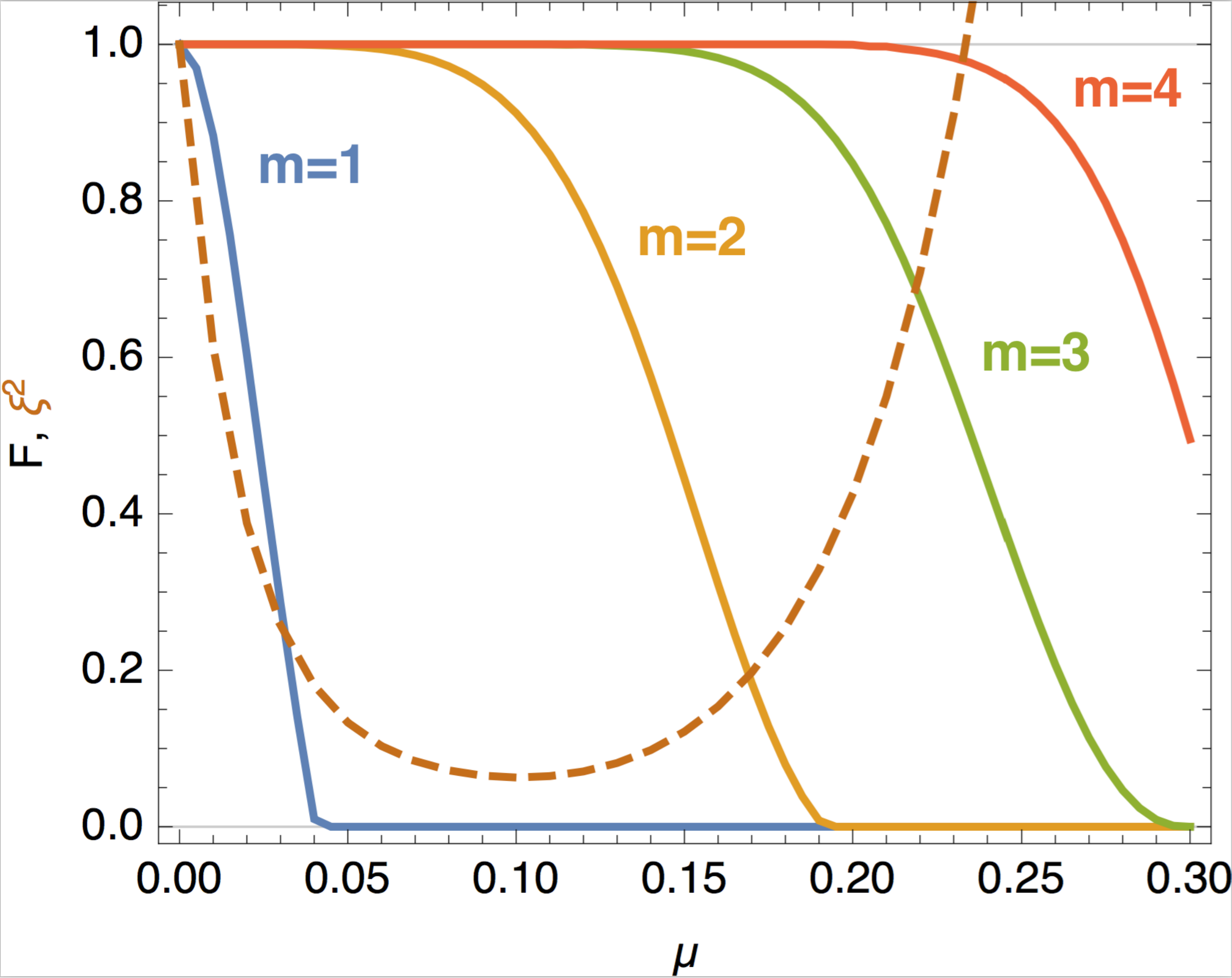}
   \caption{Fidelity bounds given by SdP \eqref{eq:gricia} for $N=100$ particles spin squeezed states $\rt$ and their $m$-RDM $\sigma$. The brown dashed line is the Wineland squeezing parameter $\xi^2$. For $\mu \gtrsim 0.10$ the state becomes significantly non-Gaussian and, therefore, it requires RDMs of larger $m$ to be faitfully represented. }\label{fig:FidSSS}  
\end{figure}

Spin squeezed states (SSS) can be prepared from a $x$-polarized coherent spin state through the action of the one-axis twisting Hamiltonian $H=\hbar\chi S_z^2$ \cite{KitagawaPRA1993}. The resulting state is conveniently parametrized by $\mu=2\chi t$, where $t$ is the interaction time. A relevant figure of merit for SSS is the Wineland spin squeezing parameter $\xi^2 = \min_\theta N \text{Var}S_\theta^2 / \vert\avg{S_x}\vert^2$, where $\theta$ is a direction in the $yz$-plane \cite{WinelandPRA1992}. If $\xi^2<1$ the state allows a quantum enhanced interferometric sequence, and entanglement between the particles is witnessed \cite{SorensenNature2001,SorensenPRL2001}. Note that for a given $N$ there is an optimal $\mu$ minimizing $\xi^2$ (see Fig.~\ref{fig:FidSSS}), which scales as $\mu_{\text{opt}}\sim 2N^{-2/3}$. 

In Fig.~\ref{fig:FidSSS} we use SdP~\eqref{eq:gricia} to bound the maximum fidelity that can be certified from knowing the full $m$-RDM of a SSS of $N=100$ particles, for different $m$ and $\mu$. We note that, while the $1$-RDM soon fails to give a fidelity $>90\%$, the $2$-RDM already suffices to obtain $F>90\%$ for $\mu\lesssim 0.14$, \ie\ for a parameter range even beyond the optimal squeezing point of $\mu_{\text{opt}}= 0.101$. This is due to the fact that within this range the SSS can be faithfully approximated by a Gaussian distribution, which is fully parameterized by its first and second moments.

The above reasoning can be extended further, to wider parameter ranges. In Fig.~\ref{fig:FidSSS} we also plot the bounds on the fidelity that can be achieved from the full $3$- and $4$-RDM, showing that, as expected, more information allows to obtain higher bounds. This reflects the fact that since the state departs from being Gaussian as $\mu$ increases, moments of order higher than second are needed to faithfully describe it.

To summarize, this example shows that a full RDM of relevant classes of states can be sufficient to obtain a nontrivial lower bound on their fidelity. However, from an experimental point of view, the full reconstruction of a RDM might still be a tedious task, as it requires to perform a considerable amount of measurements. For this reason, but also as a fundamental question, we ask whether only partial information on a RDM is still enough to obtain a nontrivial bound for the fidelity.

\section{Fidelity from partial information on the $m$-RDM}
We now address (ii), namely the effect on the fidelity bound if we impose constraints only on some of the $m$-RDM entries. Because the problem is now less constrained than the one of the SdP~\eqref{eq:gricia}, we expect to obtain a lower bound on the fidelity that is smaller or equal than before. 

Consider a set of data $D_i=\rm{Tr}[O_i \re]$, obtained from measuring on $\re$ the observables $O_i$, with $i=1,...,M$. Using this information, we can derive a lower bound on the fidelity $F(\re,\rt)$ from the following relaxation of SdP~\eqref{eq:gricia}:

\begin{equation}
  \begin{array}{llll}
  \min_{\rho} & \avg{\rho,\rt} &&\\
  \mbox{s.t.}& \rho &\succeq & 0\\
  &\langle A^{\boldsymbol \alpha}_{\boldsymbol{\beta}}, \rho \rangle &=& \sigma^{\boldsymbol{\alpha}}_{\boldsymbol{\beta}} \qquad \forall \boldsymbol{\alpha}, \boldsymbol{\beta} \vdash m \\
  &\langle O_i, \sigma \rangle &=& D_i \qquad \forall 1 \leq i \leq M \;.
 \end{array}
 \label{eq:burroesalvia}
\end{equation}

In addition, SdP~\eqref{eq:burroesalvia} allows to find \textit{a priori} what are the most effective operators to be measured in order to maximize the lower bound on the fidelity. Since in practice not all measurements have the same complexity in terms of their experimental realization, we would first like to know the one of easier implementation. To this end, we need to formulate a criterion for ``experimental complexity'' of a measurement. Clearly being platform-dependent, defining such a figure of merit is not straightforward.
Nevertheless, to workout a concrete example, for a large variety of experiments we can say that: requiring an additional measurement of the same order than the others is less demanding than introducing a new measurement of higher order. The reason behind this statement originates from the fact that estimating high order moments requires an exponentially larger number of experimental repetitions \cite{LuoJPG2012,Huang2020}. Having formulated this criterion, we proceed in finding a list of operators ordered in increasing levels of complexity.

We illustrate the procedure for collective spin measurements. First, let us remember that by exploiting the symmetry of the state, expectation values of the $k$-moment of a collective operator can be expressed as a $k$-body correlator in the $m$-RDM space (where $m\geq k$). Therefore, the first moment of the collective spin operator along direction $\vecu$, $S_{\vecu} = \sum_i \sigma_\vecu^{(i)}/2$, can be written in the $m$-RDM space as $O_1({\vecu}) = N \sigma_\vecu^{(1)} /2 $. Similarly, the second moment $S_\vecu^2$ can be written as $O_2({\vecu}) = \left( N \mathbbm{1} + N(N-1) \sigma_\vecu^{(1)}\sigma_\vecu^{(2)} \right)/4$, and so on. Then, after parameterizing the direction $\vecu$ in spherical coordinates $(\theta, \phi)$, we run an optimization over SdP~\eqref{eq:burroesalvia}, namely:
\begin{equation}
  \begin{array}{llll}
  \max_{\theta,\phi} \min_{\rho} & \avg{\rho,\rt} &&\\
  \mbox{s.t.}& \rho &\succeq & 0\\
  &\langle A^{\boldsymbol \alpha}_{\boldsymbol{\beta}}, \rho \rangle &=& \sigma^{\boldsymbol{\alpha}}_{\boldsymbol{\beta}} \qquad \forall \boldsymbol{\alpha}, \boldsymbol{\beta} \vdash m \\
  &\langle O_1(\theta,\phi), \sigma \rangle &=& \langle S_1(\theta,\phi), \rt \rangle.
 \end{array}
 \label{eq:ragu}
\end{equation}
The result of SdP ~\eqref{eq:ragu} are the optimal angles $(\theta^\ast,\phi^\ast)$ which define the first operator of our list, and the optimal fidelity bound attainable from its measurement. Then, to search for the next operator in the list we follow our complexity criterion, which in this case gives priority to an operator of the same order of $S_1$, but along a different direction. Therefore, we fix as a constraint in SdP~\eqref{eq:ragu} the expectation value of $S_1(\theta^\ast,\phi^\ast)$, and introduce the additional constraint given by measuring $S_2(\theta,\phi)$ which we again optimize over the two angles. If the result is a new measurement increasing the bound, we fix it as a constraint and repeat the same procedure, otherwise we discard the last measurement and add an additional measurement of higher order. We keep following this strategy until we reach the highest moments allowed to be measured experimentally.

To illustrate the procedure just presented with a relevant example, we consider the case of $N=100$ qubits in a spin squeezed state with $\xi^2=-5.89\;\text{dB}$ ($\mu=0.03$). The first optimization (Eq.~\eqref{eq:ragu}) gives as optimal measurement operator $S_x$, \ie the polarization of the state, from which we can obtain a lower bound on the fidelity of at most  $0.285$. We note that this is already the best lower bound that can be concluded from the knowledge of the full $1$-RDM and, therefore, additional first moments will not give any improvement (see Fig.~\ref{fig:FidSSS}, blue line). We proceed by adding one second moment, and we find that the optimal measurement $S_{\vect{n}}^2$ must be along the squeezing direction $\vect{n}$. This allows to obtain a lower bound on the fidelity of $0.999$, which is close to the bound of $1$ (up to preset numerical precision) that we expect from the knowledge of the full $2$-RDM (see Fig.~\ref{fig:FidSSS}, orange line).

To summarize, this example shows that by measuring only the first and the second moment of a collective observable it is possible to put a lower bound on the fidelity of a realistic and relevant state that is $>99\%$. Similarly, we analyzed also Dicke states, which constitute an other important class of states, and observe that their $2$-RDM is sufficient to attain unit fidelity. These exceptionally high values are reached in the ideal situation where expectation values are known precisely, and if the state is taken to be symmetric. In a practical situation, however, the measurement of physical observables will be inevitably affected by noise, and the symmetry of the state prepared experimentally is not necessarily granted. For this reason, we now analyze in detail these two limitations, and show how they can be circumvented to still obtain a non-trivial lower bound on the fidelity by using our method.


\section{Extension to noisy data}
Up to now we have taken as SdP constraints the ideal data $D_i=\rm{Tr}[O_i \re]$. In all practical situations, however, noise unavoidably contaminates measurement results and, if not properly taken into account, it could result in SdPs that are unfeasible. For this reason, we express the actual experimental data $D_i^e$ as the sum of the ``ideal'' result $D_i$ and of an error $\epsilon_i$ attributed to noise, namely $D_i^e=D_i+\epsilon_i$. Then, we modify the constraints of the SdP~\eqref{eq:burroesalvia} to be
\begin{align}
\langle O_i, \sigma \rangle &= D_i^e - \epsilon_i \nonumber\\
-\delta_i \leq & \epsilon_i \leq \delta_i ,
\end{align}
where $\delta_i$ quantifies the noise in the measurement of $O_i$, for example in units of standard deviations of the corresponding probability distribution. If available, a further characterization of the experimental noise can allow for a decrease of $\delta_i$.

\section{Extension to permutationally invariant states}
Up to now we derived a lower bound for the fidelity of an arbitrary state with respect to a fully symmetric target state. To be more general, we can further extend our work to permutationally invariant (PI) states, which describe a larger class of experimentally relevant states. Although every PI state can be purified to a symmetric state \cite{PhDRenner}, here we show a simple argument on how our algorithm can be adapted to this more general case.

Let  $\tau=\lambda \rho \oplus (1-\lambda)\rho_{\perp}$ the PI state to certify, with a symmetric ($\rho$) and a not fully symmetric ($\rho_\perp$) component, and $\Pi_S$ the projector onto the symmetric subspace. Then
\begin{align}
\avg{\tau,\rt} &= \avg{\Pi_S \tau \Pi_S, \rt} + \avg{(\mathbf{I}-\Pi_S) \tau (\mathbf{I}-\Pi_S), \rt} \nonumber\\
&= \lambda \avg{\rho, \rt} \;.
\end{align}
Therefore, the fidelity $\avg{\rho, \rt}$ given by our algorithm can simply be rescaled by the overlap with the symmetric subspace $\lambda$. The latter can also be bounded from collective observables. As an example, from the relation $\avg{\vert \vec{S} \vert^2}=S(S+1)$ we can extract $S$, which is $S=N/2$ if and only if the state is fully symmetric. Because the worst-case scenario would be when all $\rho_\bot$ is supported on the spin-($N/2-1$) sector, this yields $\lambda \geq (4\avg{\vert \vec{S} \vert^2} - N(N-1))/2N$. For further details see Appendix~\ref{app:A}.
This argument can also be adapted to provide a lower bound for the fidelity with respect to a PI target state.

\section{Bounding quantum information quantities}
Our method has direct application in a number of areas in quantum information theory, as it can bound other important quantities of interest that are directly related to the fidelity.

As a first example, the fidelity can be used as a dimensionality (Schmidt number) witness (see Eq. (5) of \cite{BavarescoNatPhys2018}). Consider a pure quantum state $\ket{\Psi}=\sum_k \sqrt{\lambda_k} \ket{k_A,k_B}$, here written in its Schmidt decomposition over a two-mode Fock basis and with $\lambda_k$ in decreasing order. The fidelity provides a lower bound on the dimensionality of (mode) entanglement since for every state $\rho$ with Schmidt rank smaller or equal than $r$, $F(\rho, \ket{\Psi}\bra{\Psi}) \leq  \lambda_0 + \cdots + \lambda_r$. Therefore, the Schmidt rank of $\rho$ can be bounded by optimizing the state $\ket{\Psi}$ so that it gives the highest $r$. 

The fidelity can also be related to entanglement measures \cite{BrussJMP2002}, like the entanglement of formation \cite{WoottersPRL1998}. In this case one typically needs to compute the fidelity $F(\rho, \ket{\Psi^+}\bra{\Psi^+})$ with respect to a maximally entangled state $\ket{\Psi^+}$.

Furthermore, the quantum Fisher information $\mathcal{F}_Q$, which is a key quantity in quantum metrology, can be bounded from the relation $\vert \mathcal{F}_Q(\rho) - \mathcal{F}_Q(\sigma) \vert \leq \zeta \sqrt{1-F(\rho,\sigma)}N^2 \;,$ where $\zeta=8$ for general quantum states, or $\zeta=6$ if one of them is pure \cite{AugusiakPRA2016}. In addition, the relative entropy (Kullback-Leibler divergence) $D(\rho\vert\vert\sigma)=\text{Tr}\left[\rho \log_2 \rho \right] - \text{Tr}\left[\rho \log_2 \sigma \right]$ can also be lower bounded from the fidelity using the relation $D(\rho\vert\vert\sigma) \geq \text{Tr}\left[\rho \log_2 \rho \right] - \log_2 F(\rho,\sigma) \;$ \cite{StreltsovNJP2010}.

These ideas could be especially relevant for the detection and characterization of correlations in experiments with \eg integrated optics \cite{WangScience2018,SATWAP, AugusiakNJP2019} and split Bose-Einstein condensates \cite{FadelScience2018, LangeScience2018, KunkelScience2018}.

\section{Conclusions}
We presented a method to give a lower bound on the fidelity between two quantum states, solely from partial information. We focused on many-particle systems, and we showed that few-body marginals are typically enough to provide non-trivial lower bounds in many cases of interest. Our approach is based on a semidefinite program, which is related to the quantum marginal problem for symmetric states \cite{Aloy2020}. To give concrete examples, we analyzed the case of spin squeezed states on which only low-order moments of the collective spin operator are measured, and we showed that the achievable fidelity can be close to unity. Similar results are obtained for other relevant classes of symmetric states we tested, such as Dicke states. In addition, we presented how our method can be used to know which are the optimal measurements to be performed in order to maximize the lower bound on the fidelity while minimizing experimental efforts. To make our tools experimentally relevant, we have discussed how to take into account measurement noise into our analysis. Moreover, even if our method is best suited to symmetric states, we explained how it applies to the wider class of permutationally invariant states. To conclude, we highlighted the usefulness of the tools we presented by showing that they can bound a number of other quantum information quantities. 
As an outlook, we mention that our results could also find application in enhanced tomography techniques \cite{TothPRL2010,Huang2020}, such as shadow tomography \cite{AaronsonSTOC2018} (see also \cite{HuberJPA2018}).

\section{Acknowledgments}
We are grateful to Giuseppe Vitagliano and Marcus Huber for useful discussions. M.F. acknowledges support from the Swiss National Science Foundation. A.A. acknowledges support from the Spanish Ministry MINECO (National Plan 15 Grants No. FISICATEAMO, No. FIS2016-79508-P, No. SEVERO OCHOA No. SEV-2015-0522, FPI), European Social Fund, Fundació Cellex, Generalitat de Catalunya (AGAUR Grant No. 2017 SGR1341 and CERCA Programme), ERC AdG NOQIA, MINECO-EU QUANTERA MAQS, and the National Science Centre, Poland-Symfonia Grant No. 2016/20/W/ST4/00314. J.T. thanks the Alexander von Humboldt foundation for support.

\appendix

\section{Bounding the state symmetry}\label{app:A}

In this section we are interested in deriving a bound for the overlap of a state with the fully symmetric subspace. Apart form allowing to bound the fidelity in a wider class of states, this approach can be applied to estimate the symmetry of an experimentally prepared multipartite quantum state from collective measurements.

Let us consider a $N$-qubit permutationally invariant state $\tau$. By virtue of the Schur-Weyl duality, there exists a symmetry-adapted basis such that $\tau$ can be written in a block-diagonal form, namely
\begin{equation}
\tau = \bigoplus_{S} {{\mathbbm 1}_{g_S}}\otimes\tau_S \;,
\label{eq:lolo}
\end{equation}
where the blocks $\tau_S$, labeled by (positive) $S=N/2, N/2-1, ...$, have dimension $(2S+1)\times(2S+1)$ and appear with multiplicity
\begin{equation}
g_S = {N \choose N/2 - S} -  {N \choose N/2 - S - 1} \;.
\end{equation}

In order to exploit this block-structure when expressing expectation values of operators, let us rewrite \cref{eq:lolo} as
\begin{equation}
\tau = \sum_{S} \lambda_S \tilde{\tau}_S \;,
\end{equation}
where $\tilde{\tau}_S$ is supported on the $S$-th blocks, and $\lambda_S$ are a probability distribution. It is then natural to decompose the expectation value of an operator $O$ as the sum of expectation values for each block, namely
\begin{equation}
\mathrm{Tr}[O \tau] = \sum_{S} \lambda_S \mathrm{Tr}[O \tilde{\tau}_S] \;.
\end{equation}

At this point, note that the expectation value $\langle{O}\rangle_{\tilde{\tau}_S} \equiv \mathrm{Tr}[O \tilde{\tau}_S]$ that a collective observables can take on each block is bounded by a function of $S$. For instance, the collective spin along direction $\vect{u}$ satisfies $|\langle{S_\vect{u}}\rangle_{\tilde{\tau}_S}|\leq S$, its second moment $\langle{S_\vect{u}^2}\rangle_{\tilde{\tau}_S}\leq S(S+1)$, and so on.
Hence, the measurement of a collective operator on $\tau$ may already give a non-trivial lower bound on the overlap of the highest-spin value $S=N/2$ block, \ie the one of fully symmetric states, as we are now going to show.

Assume that the measurement of $S_{\vect{u}}$ gives value $\avg{S_{\vect{u}}}=s$. Then, a bound on the overlap of the state with the symmetric subspace, $\lambda_{N/2}$, can be obtained from the following linear program (LP)
\begin{equation}
\begin{array}{ccc}
\min_{\lambda_S} &\lambda_{N/2}&\\
&\sum_{S} \lambda_S \mathrm{Tr}[S_{\vect{u}} \tilde{\tau}_S] &= s\\
&\sum_S \lambda_S &=1\\
&\lambda_S &\geq 0
\end{array}
\label{eq:notreallyLP}
\end{equation}
Since $\mathrm{Tr}[S_{\vect{u}} \tilde{\tau}_S] \in [-S,S]$, the only possibility to obtain a value  $s \in (N/2-1, N/2)$ is that $\lambda_{N/2} > 0$.
Note that it is actually possible to compute this minimum by defining
\begin{equation}
M = \max_{S \in [0,N/2-1]} \mathrm{Tr}[S_{\vect{u}} \tilde{\tau}_S] \;.
\end{equation}
Denoting with $x$ the index for which $M$ is achieved, the convex combination yielding the worst-case fidelity with the symmetric subspace is $\lambda_{N/2} \mathrm{Tr}[S_{\vect{u}} \tilde{\tau}_{N/2}] + \lambda_{x} M = s$, from which we obtain
\begin{equation}
\lambda_{N/2} = \frac{s-\lambda_x M}{\mathrm{Tr}[S_{\vect{u}} \tilde{\tau}_{N/2}]}.
\end{equation}
Therefore, the state $\tau$ that minimizes $\lambda_{N/2}$ is such that it maximizes both $M$ and $\mathrm{Tr}[S_{\vect{u}}\tilde{\tau}_{N/2}]$, so that $M=N/2-1$ (hence $x=M$) and $\tau$ is a mixture between the fully polarized state along the direction $\vect{u}$ in the two largest spin blocks, yielding
\begin{equation}
\lambda_{N/2} = \frac{s-\lambda_{N/2-1}(N/2-1)}{N/2} \;.
\end{equation}
Therefore, from $\lambda_{N/2-1}=1-\lambda_{N/2}$, we have
\begin{equation}
\lambda_{N/2} = \frac{2s-N+2}{2} \;.
\end{equation}

Note, however, that if the value of $s$ is smaller or equal than $N/2-1$, then $\lambda_{N/2}$ can be made zero and still satisfy \cref{eq:notreallyLP} by building $\tau$ as a mixture between \eg the $(N/2-1)$ block and any other block yielding an expectation value below $s$. Hence, in this case it is not possible to certify the fidelity even with the symmetric subspace. 

A similar argument applies \eg to the second moment of the spin operator $\avg{S_\vect{u}^2}$, or to the total spin length $\avg{\vert \vec{S} \vert^2}$. In any case, we observe that a nontrivial bound for the overlap with the fully symmetric subspace arises only if some measured observable take higher values than the one allowed by blocks with $S\leq N/2-1$. This suggests that demanding measurements (in principle with single particle resolution) are required to resolve the values in the gap between $S=N/2$ and $S=N/2-1$, and therefore that concluding the symmetry of a multipartite quantum state solely from collective measurements is a challenging task. Nevertheless, in many experimentally relevant situations, such as for Bose-Einstein condensates, the symmetry of the state could be taken as an additional assumption.

\bibliography{mylib}

\end{document}